# The Internet of Things enabling communication technologies, applications and challenges: a survey

## Sihem TLILI

UT2J, CNRS-IRIT (RMESS), University of Toulouse, Toulouse, France
Dept. of Computer Sciences, Faculty of Sciences, University of Gafsa, Gafsa, Tunisia
Email: sihem.tlili@irit.fr

## Sami Mnasri

CNRS-IRIT (RMESS), University of Toulouse, Toulouse, France
Dept. of Computer Sciences, Community College, University of Tabuk, Tabuk, KSA
Email: sami.mnasri@irit.fr

## Thierry Val

UT2J, CNRS-IRIT (RMESS), University of Toulouse, Toulouse, France
Email: val@irit.fr

**Abstract:** Recently, the IoT has gained great importance, which results in the evolution of communication technologies to meet the needs of different IoT applications. In addition, several domains integrate the IoT technologies in several applications of our professional life and daily activities. However, the IoT still presents many challenges and issues. This article describes in details the emerging communication technologies used in the IoT networks and enumerates the common domains of their application. It also describes the main challenges of the IoT and its use in order to exploit its advantages at most.

## 1  Introduction

The International Telecommunications Union defined the Internet of Things as that involving far-reaching technological and social implications to form a global infrastructure of the information society. IoT uses information and communication technologies to make physical and / or virtual objects interconnect in order to provide advanced services in our professional life and daily activities. Thanks to its advantages, the IoT is increasingly integrated in many application domains.

The IoT environments contain many types of sensors that measure the physical parameters of the environment and transform them, first, into electrical signals and, then, into digital messages. Furthermore, they produce a variety of data in different formats. In fact, the success of the IoT is attributed to an efficient management of these data that should be processed and communicated in real time and in a reliable and rapid manner in certain critical and time sensitive applications. Therefore, new technologies and new approaches must be introduced and integrated to manage efficiently these amounts of data. Moreover, communication technologies are constantly evolving in order to meet the needs of the IoT and provide solutions that can be adapted to different fields of application. Thus, a wide variety of IoT applications has been recently developed and deployed to significantly affect our daily and professional lives. Added to that, the emergence of new use cases has led to the appearance of several application domains of IoT. On the other hand, despite its success, the IoT still faces many challenges and problems. The major issues are: the management of Big Data (collection, transmission and analysis), the security and confidentiality of these data and the interoperability between the used heterogeneous technologies.

This paper provides an overview of the Internet of Things (IoT) by focusing on: i) the used communications technologies, ii) the most popular application domains of the IoT and iii) the main challenges faced by these applications. The reminder of this paper is organized as follows. Section 2 lists the emerging communications technologies used in the IoT networks. It also presents a comparative study between these technologies using some criteria while specifying their common use cases. Then, Section 3 describes a variety of application domains of the IoT and their common challenges. Finally, Section 4 concludes this paper.

## 2 Communication technologies for the IoT

### 2.1 Wired technologies

According to (Dong et al., 2019) and (Alparslan et al., 2020), the wired communication technologies that are used in the IoT networks are: Ethernet, HomePlug GP, HomePNA, HomeGrid/G.hn and MoCA. These technologies are summarized in Table 1 and illustrated in the rest of this section.

- Ethernet (IEEE 802.3): It is one of the common local area network (LAN) technologies used in real life. Ethernet links several machines through twisted pair cables. The physical cabling makes this technology more secure and less vulnerable to disturbances. Table 1 contains a description of some characteristics of this technology. In (Jiabul et al., 2020), Ethernet was defined as a wired technologies that can be used in Smart Homes by providing reliable and secure transmissions. Besides, it is a low-cost technology with a high data rate (from 10 Mbps to several Gbps) (Gianfranco et al., 2018) based on IPv4 / IPv6 (Internet Protocol versions 4/6). Indeed, the latest versions of Ethernet (802.3cu-2021 and 802.3cv-2021) offer up to 400 Gb/s on single-mode fiber (IEEE et al., 2021). On the other hand, its range is limited to 100m (Jiabul et al., 2020) for each cable linking two active elements (hub and switch). Authors, in (Thompson, 2019), described the transition to the use of Ethernet in the IoT. (Bahl et al., 2016) and (Pintilie et al., 2019) detailed the cases where Ethernet is used in IoT applications. (Xuepei et al., 2019) presented a performance evaluation of the use of Ethernet in an industrial IoT environment.

- HomePlug Green PHY (GP): it is a power-line communication (PLC) technology. This technology uses the existing electrical wiring in a building to connect network devices. HomePlug employs powerline adapters that plugs into a power outlets and use Ethernet cables to be connected to other devices. It can be utilized, for example, with an existing Ethernet network to connect equipment in a remote room without using wireless technology. The most important advantage of the HomePlug GP is its economical energy consumption (Dr. Yusuf et al., 2019). Besides, it offers a data rate between 4 Mbps and 200Mbps. Its range is close to 300m (Jiabul et al., 2020). It is characterized by security, reliability, compatibility with existing infrastructure and interoperability with IEEE 1901 devices (Md. Liton, 2020). Other characteristics of this technology are represented in Table1. HomePlug GP is designed for the IoT applications, mainly in Smart Homes and smart grid applications (Dr. Yusuf et al., 2019). In addition, due to the new power management and features offered by HomePlug GP 1.1, it has been a promising alternative for green communications for electric vehicles (Pallander, 2021).

- HomePNA (Home Phone Networking Alliance): It is a technology applied in home networks by reusing telephone and coaxial cables. Indeed, HomePNA networks use phone jacks, HPNA adapters and bridges with coaxial cables and telephone wiring to connect devices. Using the existing cables does not interrupt voice or television services since different frequencies are used (Table 1). HomePNA is mainly employed in Smart Homes with a range close to 300m (Jiabul et al., 2020). HomePNA, described in Table 1, offers a data rate up to 320 Mbps (HPNA 3.1) (Gianfranco et al., 2018). 4.0 version of HomePNA includes support for the G.hn standard, which makes all connections possible regardless of the type of wiring.

- HomeGrid/G.hn: It is a technology that utilizes the existing types of domestic cabling (electrical cabling (PLC), coaxial cable and telephone lines) and plastic optical fibers (Phatrakit et al., 2019), which makes it very cost effective. Indeed, a G.hn network uses adapters to connect its devices. It can be used in Smart Homes (Jiabul et al., 2020) and offices as well as in industrial applications (Zeng, 2020). It is more beneficial than HomePNA as it provides a higher data rate (1 Gbps over power lines and 1.7 Gbps over phone lines and coaxial cables) with more robust performance. However, its range is limited to 100m (Phatrakit et al., 2019). Other characteristics of G.hn are represented in Table1.

**Table 1** The characteristics of Ethernet, HomePlug PG, HomePNA, HomeGrid/G.hn and MoCA

| Technology | Ethernet | HomePlug PG | HomePNA | HomeGrid/G.hn | MoCA |
|---|---|---|---|---|---|
| *Used infrastructure* | Ethernet cables and jacks | Electrical wiring and power outlets | Telephone and coaxial cables, phone jacks | Power lines, coaxial cables and phone lines plastic optical fibers | Coaxial cables and jacks |
| *Frequencies* | - | 24 - 500kHz | 12 - 44MHz | Power lines: 100MHz Coaxial cables: 200MHz Phone Lines: 200MHz | 500 - 1650MHz |
| *Range* | 100 m | 300m | 300m | 100m | 90m |
| *Data rate* | 400 Gb/s | between 4 Mbps and 200Mbps | 320Mbps | 1 Gbps over power lines 1.7Gbps over phone lines and coaxial cables | 10 Gbps |
| *Latest version* | 802.3cv-2021 | HomePlug GP 1.1 | HomePNA 4.0 | G.hn Wave-2 | MoCA 3.0 |
| *Common Use Cases* | smart home, smart offices | smart homes and offices, smart grid applications, electric vehicles | smart homes and offices | smart homes and offices, industrial applications | smart homes, buildings, hospitals, hotels and schools |

- MoCA (Multimedia over Coax Alliance): It is a technology that uses coaxial cables (Jiabul et al., 2020) and can be employed in Smart Homes with a very high reliability, security and data rate (Dragos et al., 2018). Its characteristics are described in Table 1. Its latest version MoCA 3.0 offers a data rate of 10 Gbps. Its range is close to 90m. On the other side, MoCA network can utilize 16 adapters simultaneously, which makes it useable in buildings, hospitals, hotels and schools. Its combination with Wi-Fi repeaters increases the Wi-Fi range in a reliable way (without loss of data rate).

### 2.2 Wireless technologies

The most widely used wireless communication technologies in the IoT networks are described in this section. Tables 2, 3 and 4 detail their characteristics. Data illustrated in this section and these tables show the information presented in (Dong et al., 2019), (Wajih et al., 2020), (Naik, 2018), (Bharat et al., 2020), (Saleem et al., 2019), (Joel et al., 2019), (Evgeny et al., 2018), (Kais et al., 2019), (Ayoub et al., 2019), (Joseph et al., 2018), (Franck et al., 2019), (Ahmad et al., 2020) and (Saurabh et al., 2020).

- RFID (Radio Frequency Identification): It is a communication technology whose main objective is to store small data and, then, retrieve it remotely. It is used to transmit small amounts of data between an RFID tag and a reader (Figure 1(a)). RFID tag is a small device that plays the role of a transponder. It contains data that can be retrieved by being transmitted remotely after receiving a signal from the reader. RFID reader is a transceiver that generates radio waves through antenna to get information from the RFID tags. It also transfers the received data to a network.

Table 2 exposes the RFID characteristics. The range of this technology is short. In fact, the RFID is commonly used in retail applications and asset as well as in inventory tracking. It can also be used in an industrial environment to identify, in a simple and an efficient way, some measurements and information (physical location, offset measurements, identification and control of raw materials, etc.).

There are three versions of the RFID technology: 1) a passive version: It is the most economical version where the tags use the energy generated by the waves of the RFID readers since they are not equipped with an internal battery; 2) an active version where the tags are powered by their own energy sources; 3) a semi-passive version in which the tags utilize their own energy sources and the energy generated by the RFID readers.

**Figure 1** RFID and NFC Systems Architectures

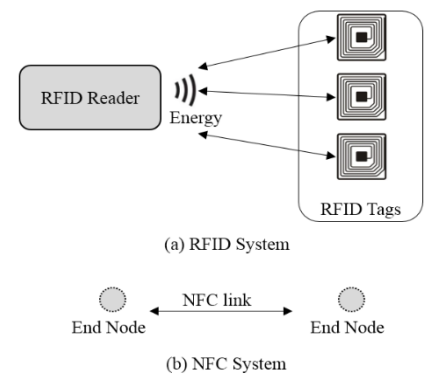

(a) RFID System

(b) NFC System

- NFC (Near Field Communication): It is a communication technology in which two devices can exchange information (Figure 1(b)) by a wireless link at a short distance equal to 4 centimeters. Its basic structure components are the same as that of RFID. In fact, an end node in an NFC system can be: i) an NFC tag: a transponder that stocks data and sends it after receiving a signal from a reader, ii) an NFC Reader: a transceiver that reads data from tags, or iii) a device that combines both an NFC reader and an NFC tag.

The proximity between the devices that communicate with the NFC reduces the risk of hacking. For this reason, the NFC is commonly used in electronic payments and tickets. This technology is also supported by several devices that use Android and IOS systems. Table 2 contains a detailed description of NFC.

- BLE (Bluetooth Low Energy): It is a short-range wireless network technology employed in Personal Area Networks (PAN). BLE is a version of Bluetooth standard designed to transmit / receive small amounts of data while consuming extremely low amounts of energy. Its possible architectures and end nodes (devices equipped with BLE technology) are represented in Figure 2. The first possible architecture, called Point-to-Point (Figure 2(a)), creates direct communications between two BLE nodes. In a star architecture (Figure 2(b)), the BLE network includes a number of end nodes connected to the BLE Central Master. It can represent a bridge to the outside of the network. In a mesh architecture (Figure 2(c)), the network contains full-interconnected BLE end nodes.

**Figure 2** Possible BLE Architectures

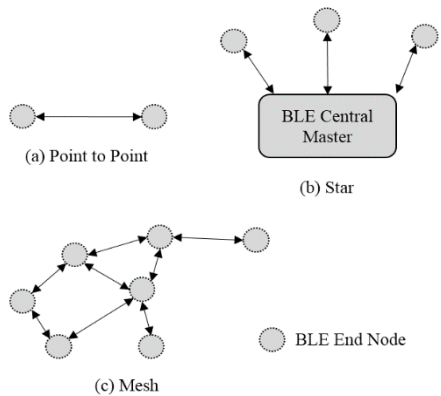

In fact, the BLE is supported by several operating systems (Linux, Windows, macOS, Android and IOS). Thanks to the simplicity of its implementation, it is widely used to transmit sensor data. This technology is essentially applied in Smart Home, smart wearable devices, smart health and retail. Indeed, it makes it possible to transmit data from portable devices (fitness, medical, Smart Home, etc.) to smartphones. Table 2 presents the characteristics of BLE.

- ZigBee (IEEE802.15.4): It is a medium range technology designed for economical wireless networks with a low power consumption. This technology can put objects in a dormant state if necessary. ZigBee is used mainly in Smart Homes since it connects the nodes that are nearby the medium range. Its architecture is represented in Figure 3. Its main components are: i) ZigBee Coordinator, ii) ZigBee end nodes (which are devices equipped with ZigBee technology) and iii) ZigBee Routers that act as intermediates between end devices and the coordinator. The ZigBee Coordinator is the only device that can form and start the network. Other devices can join the network after sending requests. In addition, the coordinator may function as a bridge to other networks. Zigbee network must have one coordinator.

**Figure 3** ZigBee Architecture

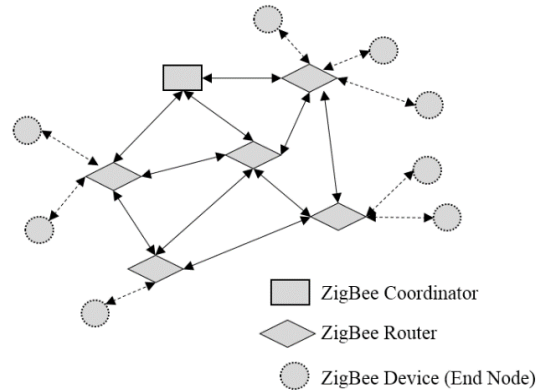

ZigBee offers robust and highly secure operations and facilitates the addition or removal of network objects. However, adding a new device requires checking its compatibility with the current components of the network. As the ZigBee is available in an open source, it can be modified to be adapted to particular devices although the latter cannot communicate with other network components. Another disadvantage of the ZigBee is the signal disturbances since it uses a frequency that includes other technologies and equipment in the same environment (e.g. WiFi, Bluetooth, microwaves, etc.). A detailed description of ZigBee is represented in Table 2. The Zigbee Alliance announced its most recent IP standard, Matter, to simplify and harmonize the IoT. It is a secure connectivity standard that solves the problem of interoperability in Smart Homes. Indeed, the objective of this standard consists in making a greater number of connected objects interoperable by creating an IP-based interoperability layer at the application layer.

- Z-Wave: it is a communication technology designed primarily to automate home. Indeed, it can make objects communicate effectively with one another with low energy consumption. Moreover, Z-Wave facilitates the addition of a new compatible device since it adds another mesh to the network. Unlike the ZigBee, the Z-Wave is less disturbed by other devices since it uses a frequency of 868 MHz. The Z-Wave architecture is shown in Figure 4. It contains two types of components: i) Z-Wave slaves (which are the end devices equipped with ZigBee technology using the network to communicate) and ii) Z-Wave controller (which is the component that organizes the network and controls all other devices to work together).

Z-Wave is represented in Table 2.

**Table 2** Comparison between the characteristics of BLE, NFC, RFID, Z-Wave and Zigbee

| Technology | BLE | NFC | RFID | Z-Wave | ZigBee |
|---|---|---|---|---|---|
| Category | WPAN | Near Field | Near Field | WPAN | WPAN |
| Frequency | 2.4 Ghz | 13.56 MHz | 13.56 MHz | 868 MHz | 868 MHz (EU), 2.4 GHz (global) |
| Power Consumption | Very low | Very low | Very low | Low | Low |
| Range | From 50 to 150 m (70m typical) | 10 cm (4cm is the optimal range) | 100m (active tags), 25m (passive tags) | 100m | From 10 to 100m |
| Data rate | 1 Mb/s | 100-420 kbps | 500 Kbps | 100 Kbps | 20 kbps (868 MHz), 250 kbps (2.4 GHz) |
| Maximum Transmission Unit | 251 bytes | | 16-64 Kbps | 255 bits | 100 bytes |
| Bidirectional Communication | yes | yes | yes | yes | Yes |
| Topology | Point to point, star and mesh (Bluetooth MESH) | Point to point | Point to point, star | Mesh | Mesh |
| Scalability (devices) | 5917, 50 (mesh) | 2 | limited | 232 | 65 000 |
| Mobility Support | yes | yes | yes | yes | Yes |
| Direct Internet Access | no | no | no | no | No |
| Common Use Cases | Smart home, smart wearable devices, smart health, retail | Electronic payments and tickets | Retail applications, asset and inventory tracking | Medium-range IoT applications whose nodes are nearby. For example: home automation | Medium-range IoT applications whose nodes are nearby. For example: home automation |

**Figure 4** Z-Wave Architecture

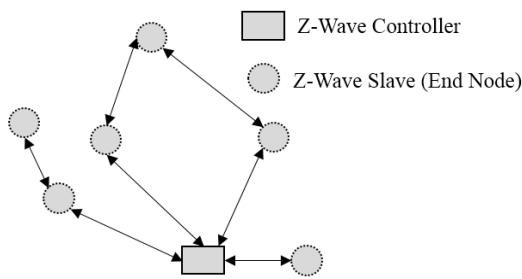

- WiFi (Wireless Fidelity): It is widely used since it makes objects connect to each other or directly to the Internet. Table 3 shows the characteristics of WiFi. Figure 5 exposes the architecture of a use case of WiFi in an IoT network. The components represented in the figure are: i) End nodes (WiFi devices that ensure communication); ii) WiFi Mesh Routers (the routers that transmit data between end nodes and the gateway to another network (Internet). These routers can communicate with each other in a mesh topology); and iii) WiFi Mesh Router with Gateway (a router that acts also as a gateway to other networks). WiFi is efficiently used to transfer files thanks to its ability to handle large amounts of data. It is mainly characterized by its fast transfer speed. However, it has limits in terms of range and power consumption. Its best use cases are Smart Home and smart retail.

**Figure 5** The architecture of the WiFi in an IoT network

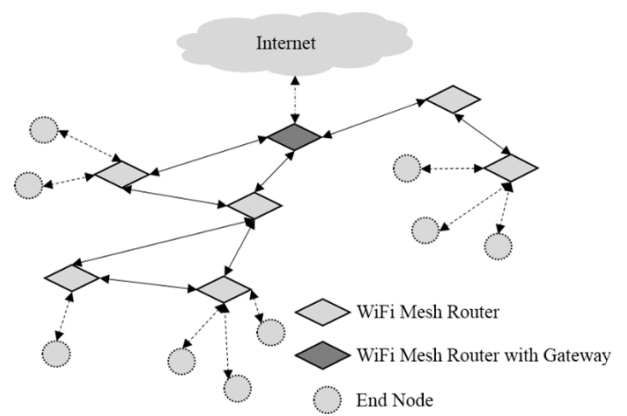

- LoRaWAN: It is a LPWAN (Low-Power Wide Area Networks) technology that provides long-range communications with low power consumption. LPWAN technologies are adequately used in applications that do not require high data rate and are not time sensitive. They are generally employed in large industrial and commercial environments. They are also utilized in smart city, building and agriculture applications. LoRaWAN is a technology that transfers data from terminal nodes to the central server, via gateways, by a single hop. It is developed to support LoRa over the Internet. Its architecture is shown in Figure 6 where the typical system architecture of a LPWAN is represented. Its main components are: i) end nodes (the end devices that communicate data); ii) network server (the central point of the network that aggregates data and acts as an intermediate with application servers); iii) LPWAN gateways (the devices that communicate data between terminal nodes and the central server); and iv) application servers (the servers that run applications using the network data). LoRaWAN is represented in Table 4.

**Table 3** Comparison between the characteristics of the different WiFi versions

| WiFi version | 802.11 Standard | 802.11a | 802.11b | 802.11g (WiFi 3) | 802.11n (WiFi 4) | 802.11ac (WiFi 5) | 802.11ad (Millimeter wave) | 802.11 ah (WiFi Halow) | 802.11ax (WiFi 6) |
|---|---|---|---|---|---|---|---|---|---|
| Category | WLAN | WLAN | WLAN | WLAN | WLAN | WLAN | WLAN | WLAN | WLAN |
| Frequency | 2.4 GHz | 5 GHz | 2.4 GHz | 2.4 GHz | 2.4 GHz / 5 GHz | 5 GHz | 60 GHz | 0.9 GHz | 2.4 GHz / 5 GHz |
| Power Consumption | High | High | High | High | High | High | High | Low | Low |
| Range | 20m (Indoor), 100m (Outdoor) | 100m | 100m | 100m | 100m | 100m | 10m | 1000m | 100m |
| Data rate | 1 Mbps-2 Mbps | 54 Mbps | 11 Mbps | 54 Mbps | 600 Mbps | 6.9 Gbps | 7 Gbps | 347 Mbps | ≈ 9.6 Gbps |
| Maximum Transmission Unit | 2304 bytes | 2304 bytes | 2304 bytes | 2304 bytes | 2304 bytes | 2304 bytes | 2304 bytes | 100 bytes | 100 bytes |
| Bidirectional Communication | colspan Yes | | | | | | | | |
| Topology | Star and Mesh | | | | | | | | |
| Scalability (devices) | - | 2007 | 2007 | 2007 | 2007 | 2007 | 2007 | 8191 | A higher number |
| Mobility Support | Yes | | | | | | | | |
| Direct Internet Access | Yes | | | | | | | | |
| Common Use Cases | smart home and smart retail | | | | | | | long distances without consuming a lot of energy | |

- Sigfox: It is an LPWAN technology that ensures the transmission of small amounts of data. Initially, Sigfox supports communications from terminal nodes to base stations. However, it is not efficient for communications in the opposite direction (from base stations to terminal nodes). Later versions support two-way communication between terminal nodes and base stations. The architecture of Sigfox is the same as that of a general LPWAN (Figure 6). Table 4 contains a detailed description of the characteristics of Sigfox.

**Figure 6** Typical LPWAN System Architecture

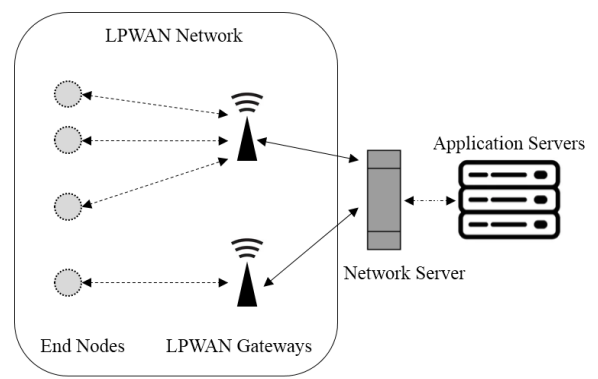

- NB-IoT (Narrowband IoT): It is an LPWAN cellular technology designed specifically for IoT networks. It efficiently connects the IoT devices over mobile networks by processing small amounts of bidirectional data in a secure and reliable manner. It is optimized to exchange small amounts of data in networks using a large number of devices and over long periods. NB-IoT architecture is depicted in Figure 7. Its main components are: i) end nodes (the end devices that generate data to be sent to the base stations and can receive data from them); ii) evolved base stations (the base stations that control the end devices, communicate with them and relay data to IoT Evolved Packet Core); iii) IoT Evolved Packet Core (EPC) (the component that connects the network with IoT platform and application server); (iv) application server which is the server that hosts and runs applications; and v) IoT platform (a set of components that facilitate the interactions between the IoT EPCs and the application server and orchestrates the data movements). The characteristics of NB-IoT are illustrated in Table 4.

**Figure 7**  NB-IoT Architecture

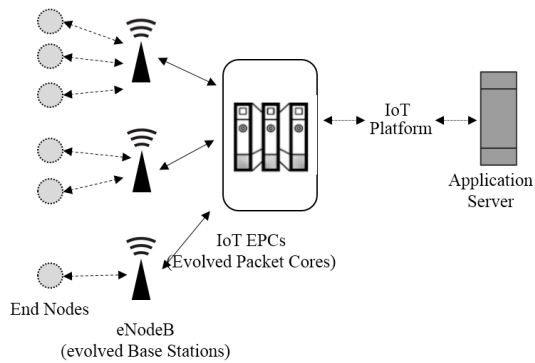

- NB-Fi (Narrowband fidelity): It is an LPWAN technology designed for bidirectional communications between machines (M2M). It is characterized by low power consumption. NBi-Fi provides robust and reliable communications and uses artificial intelligence techniques to improve its performance and to avoid interference. NB-Fi network has the architecture of a LPWAN exposed in Figure 6. The characteristics of NB-Fi are listed in Table 4.

- DASH7: It is an extension of the RFID. It offers communications with low latency and high flexibility. DASH7 provides multi-hop communications and supports object mobility up to 2 km range. The DASH7 architecture, which is similar to LoRaWAN, is represented in Figure 6. Table 4 includes the characteristics of DASH7.

- Cellular (3G/4G/5G): These technologies provide high data rate and reliable communications over long distances. Moreover, they require very high operating costs and energy requirements. An example of the architecture based on 5G used in the IoT is presented in Figure 8. It is composed of : i) end nodes (the end devices equipped with 5G technology; ii) Gateways (the devices that connect the 5G network to other IoT networks); and iii) 5G Base Station which is the component that connects other devices (end nodes and gateways) to the Internet. Table 4 enumerates the characteristics of the 5G. Although the 5G offers several advantages (e.g. considerable improvement of the data rate and the reduction of latency). The next version (the 6G) was developed to introduce a communication standard of new generation with more advantages and enhanced capacities.

**Figure 8**  The architecture of an example of using 5G for the IoT

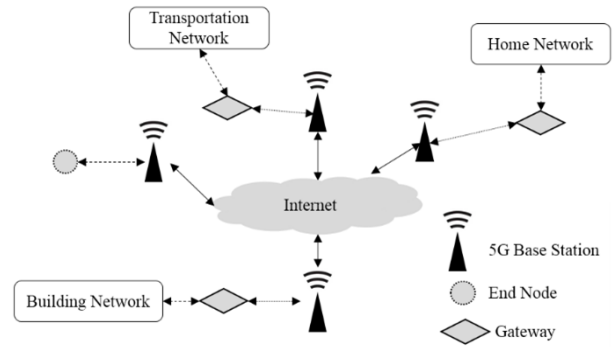

- LiFi (Light Fidelity): It is a visible light (VLC) wireless communication technology that uses visible light to transmit data. LiFi is a promising technology for the IoT networks since it makes the latter meet their requirements (a high data rate, high power efficiency, high security, no interference, more reliability, low cost, etc.) even with a large number of connected devices (Ahmad et al., 2020) (Qusay et al., 2020). In fact, it avoids the negative impact of radio frequency (RF) on the human health and reduces carbon emissions. In addition, it can be used simultaneously for communication and lighting. However, this technology suffers from limited range and high vulnerability to interruptions (light cannot pass through obstacles). Therefore, it cannot be used in environments where there are other light sources. For this reason, the LiFi technology can be used only in specific fields where it coexists with other technologies since it cannot completely replace them (Ahmad et al., 2020). Figure 9 presents a scenario that combines both the WiFi and the LiFi technologies. This network is composed of: i) end nodes (the end devices that communicate data); ii) Relay/Amplification devices (the components that relay data between end devices and the LiFi Access Point and can amplify the signals while transmitting the data); iii) LiFi Access Point (the device that aggregates data from LiFi devices and transmits it to WiFi Access Point); and iv) WiFi Access Point (the device that connects other devices to a public network). The characteristics of LiFi are represented in Table 4.

**Figure 9**  A topology that combines LiFi and WiFi

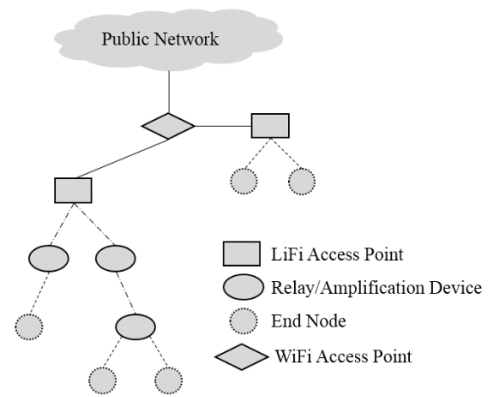

**Table 4** Comparison between the characteristics of 5G, LiFi, LoRa, Dash7, Sigfox, NB-IoT and NB-Fi

| Technology | Cellular (5G) | LiFi | LoRa | Dash7 | Sigfox | NB-IoT | NB-Fi |
|---|---|---|---|---|---|---|---|
| Category | WWAN | WLAN | | | LPWAN | | |
| Frequency | 25-39 GHz | 10000 times of WiFi | 125 KHz/ 250KHz | 433 MHz | 100Hz | 200 kHz | 868 MHz et 915 MHz |
| Power Consumption | 2 times less lower than 4G | Low | low | low | low | low | low |
| Range | from 200m to 400m | Based on light intensity (<=10m) | 2-5 km (urban areas), 20 km (rural areas) | Up to 5km | 30-50 km (rural areas), 3-10 km (urban areas) | 1 km (urban areas), 10 km (rural areas) | 16km (urban areas), 50km (rural areas) |
| Data rate | up to 36 Gbps | About 100Gbps1Tbps | from 300bps to 50 kbps | up to 167 kbps | 100 bps | 200kbps (downlink), 20kbps (uplink) | 100bps |
| Maximum Transmission Unit | 1420 bytes | - | 250 bytes | 256 bytes | ≈ 12 bytes | 1600 bytes | - |
| Bidirectional Communication | yes | yes | yes | yes | yes | yes | yes |
| Topology | star | Point to Point and Star | star, a star of stars, mesh | star, tree (if there are subcontrollers) | star | star | star |
| Scalability (devices) | 1000000 by km$^2$ | unlimited number of devices present in the visibility zone | >10000 | - | >10000 | ≈ 55000 | up to 2 million nodes for a base station |
| Mobility Support | yes | yes | yes | yes (node mobility up to a range of 2km) | yes | yes (Complex) | - |
| Direct Internet Access | yes | no | yes | yes | yes | yes | yes |
| Common Use Cases | Smart health | Smart cities, Healthcare applications, Underwater applications | Applications that do not require high data rate but require a large range and are not time sensitive. Examples: Industrial IoT, smart city, smart building, smart agriculture | | | | |

## 2.3 Wired or wireless

Since each communication technology has its own strengths and weaknesses in terms of several criteria, there is not a universal communication solution suitable for all the use cases and all the IoT applications. Every user must select the right technology according to the characteristics of the corresponding use case. The first step is to choose between the two main categories of communications technologies: wired and wireless.

Table 5 presents the advantages and disadvantages of each of these categories and compare them according to some criteria. This comparison demonstrates that wired technologies are characterized by lower power consumption and ensure faster and more secure transmissions, compared to the wireless technologies. Indeed, the use of wires makes it possible to create connections with high data rate and more reliability. However, wireless technologies reduce the data rate and can cause reliability problems in real time because of the errors caused by interference, noise and collisions. In addition, they are highly sensitive to intrusion and hacking, which makes wired technologies more secure. Despite these disadvantages, wireless technologies have some advantages. For instance, they create networks with longer ranges, compared to the wired technologies. These wireless networks are also more scalable and more flexible since new objects and mobile devices can be easily added and a greater number of connected objects can be supported using these technologies. In addition, they are less expensive and time-saving since they do not require the installation of many equipment and because of their easy remote maintenance. Moreover, without wires, the risk of failures and dangers can be minimized since their use can sometimes be dangerous, especially in high-temperature environments. All these advantages make wireless technologies a good choice for IoT networks. According to (Dong et al., 2019), wireless communication technologies were behind the development of the IoT since they improve its system flexibility and scalability by providing long-range and low-power communications.

**Table 5** Comparison between wired and wireless communications technologies ("+" means "more advantageous" and "-" means "less advantageous")

| Criterion | Wired technologies | Wireless technologies |
|---|---|---|
| Range | - | + |
| Data rate | + | - |
| Management of mobile nodes | - | + |
| Power consumption | + | - |
| Reliability | + | - |
| Security | + | - |
| Costs | - | + |
| Scalability | - | + |
| Flexibility | - | + |
| Speed of installation | - | + |
| Ease of maintenance and reconfiguration | - | + |
| Risk of failures | - | + |
| Danger of use | - | + |

From the above-presented comparisons, we can deduce that wireless communications technologies are better suited for IoT environment. However, despite their importance, these technologies cause some problems. Indeed, due to the error resulting from interference, data to be transmitted can be lost. In addition, the transmission errors may lead to variable transmission delays, which can negatively affect the quality of the services provided by some applications. Moreover, the use of a wireless medium may limit the number of devices that share it depending on its bandwidth. Besides, compared to wired technologies, the data rate offered by wireless communication technologies is limited. For this reason, the deployed solutions must manage data efficiently and reduce these undesirable effects (losses and waste) by optimizing the use of the available resources.

## 3  Application domains of the IoT

Thanks to its benefits, the IoT becomes more integrated into different fields of application. In fact, its implementation is growing exponentially although it faces many challenges. In this section, we will present the most popular application domains of the IoT and their most important challenges.

- Smart City: is used in many domains such as public safety, greening the environment, smart lighting, traffic management, private and public transport traffic, efficient management of waste, etc. Thanks to this development, cities become smarter by offering more high-quality services and reducing their expenses (Dr. Yusuf et al., 2019).

- Smart Home: Authors, in (Dr. Yusuf et al., 2019), defined Smart Home as a residence equipped with IoT technologies. Its equipment is connected to the Internet in order to control continuously and remotely the various activities of the residents based on data collected from the deployed sensors. Indeed, it offers a better quality of life with more safety and comfort.

- Smart Building: The IoT can make buildings smart by benefiting from its solutions in terms of monitoring and management. Moreover, the integration of the IoT allows reducing the operating and maintenance costs of the buildings to better satisfy the occupants by guaranteeing their safety and minimizing the environmental impact of the buildings in terms of energy, water consumption and waste production.

- The Educational Internet of Things (EIoT): It is used to apply IoT technologies in education and learning. In fact, carrying out the learning process via the IoT makes it possible to build intelligent educational environments (intelligent technologies, intelligent objects and intelligent pedagogy) to make it more productive and more efficient.

- The Agricultural Internet of Things: Recently, the agriculture has become more industrialized to improve production quality and quantity by integrating the Agricultural Internet of Things (AIoT) that consists in using the IoT technologies to modernize and enhance the various application fields of the agricultural sector. AIoT can improve the agricultural productivity and minimize the human intervention and workforce.

- The Industrial Internet of Things (IIoT): It consists in using the IoT technologies in an industrial environment. It is employed to equip the industry with sensors and actuators in order to obtain automated intelligent objects that detect and communicate real-time events (Wazir et al., 2019), receive orders and act by applying the recommended actions.

- The Medical Internet of things (MIoT): It is based on integrating the IoT technologies in the healthcare sector. In this context, (Fadi et al., 2020) defined the MIoT network as a network of people and medical devices that interchange health data through wireless communication. Therefore, the MIoT improves the quality of medical treatment, reduces healthcare costs and creates more efficient medical systems (Mohan, 2020) (Fadi et al., 2020).

- The Energy Internet of Things (EIoT): It applies the IoT technologies in energy distribution networks. In

order to automate the energy resources and use them in a smart and efficient way, the IoT facilitates the collection, exchange and use of data. It accelerates the integration of renewable energy. In fact, an efficient localization method should be applied in this type of networks. In this context, (Vey et al., 2020) offered a distributed algorithm for wireless nodes localization in sparse wireless networks. The simulation results revealed its improved performance in terms of localization accuracy.

**Table 6** IoT Application Domains: target environments, target customers and the most important objectives

| Application | Target Environments | Target Customers | Important objectives |
|---|---|---|---|
| Smart City | Urban areas<br>City and local government agencies, public and private local utilities and quasi government/private sector consortiums | Citizens | Improves the quality of the citizens' life<br>Offers good-quality and more comfortable services<br>Maintains the sustainability of services<br>Reduces the operating expenses<br>Helps to deal with congestion and energy waste |
| Smart Home | Residences | Homeowners<br>Inhabitants | Satisfies the individual's preferences and needs<br>Provides convenience in daily activities<br>Space-saving, security and energy efficiency<br>Reduces human involvement |
| Smart Building | Buildings, hospitals, hotels, schools, etc. | Residents | Satisfies the occupants and guarantee their safety<br>Reduces the operating and maintenance costs<br>Minimizes the environmental impact of the buildings in terms of energy and water consumption and waste production. |
| Educational Internet of Things (EIoT) | Real and virtual educational environments | Learners<br>Teachers | Makes educational environments intelligent, more productive and more efficient<br>Provides collaborative and personalized self-learning and learning opportunities |
| Agricultural Internet of things (AIoT) | Agricultural environments | Customers and owners of agricultural environments | Improves production quality and quantity<br>Modernizes and improves the various application fields of the agricultural sector.<br>Minimizes the human intervention and workforce efforts |
| Industrial Internet of Things (IIoT) | Industrial environments | Workers and owners and of industrial environments | Achieves a high operational efficiency and increases productivity<br>Ensures better management and control of the equipment and the industrial processes<br>Makes the industrial operations fully autonomous |
| Medical Internet of things (MIoT) | Healthcare environments | Patients<br>Workers in healthcare environments | Provides personalized and targeted medicine<br>Improves the quality of medical treatment<br>Reduces healthcare costs<br>Creates more profitable systems |
| Energy Internet of Things (EIoT) | Energy distribution networks | Customers and providers of energy distribution networks | Automates the energy resources and uses them in a smart and efficient way<br>Optimizes energy consumption<br>Incorporates renewable energies into distribution networks |
| Internet of Battlefield Things (IoBT) | Military environments | Soldiers | Saves soldiers' lives and provides autonomous war environment<br>Implements human monitoring of the war environment<br>Improves the operational efficiency of military applications |
| Commercial Internet of Things (CIoT) | Real and virtual Commercial environments | Customers, businesses, sellers and manufacturers | Extends e-commerce and mobile commerce<br>Makes the purchasing process automatically driven by algorithms and minimises human intervention<br>Recognizes the customer's needs, improves his/her proposals and tracks his/her consumption/usage |
| Social Internet of things (SIoT) | Any IoT environment | All users of IoT networks | Uses object relationships to discover objects, the information they provide and their services<br>Improves the performance and functionality of the IoT objects |
| Green Internet of Things (GIoT) | Any IoT environment | All users of IoT networks | Reduces energy consumption and minimises the emissions of toxic pollutants<br>Involves environmental conservation activities |

- The Internet of Battlefield Things (IoBT): It consists in applying the IoT in the military domain. In other words, devices used in the world of defence and military battles become intelligent as they can communicate with each other. As a result, they become more efficient as they improve the operational efficiency of military applications.
- The Commercial Internet of Things: (Henner et al., 2020) defined the CIoT as the use of IoT devices to purchase products and services online in order to give new opportunities to retail customers. In fact, the IoT affects all stages of the buying process during the interactions between the client and the seller or manufacturer.
- The Social Internet of Things (SIoT): It is the combination of social networks with the IoT technologies. In fact, the SIoT networks include intelligent objects considered social as they mimic the human behavior in social networks. Therefore, these objects can create their social networks (Mozhgan et al., 2020) by making relations and, thus, constructing social structure.
- The Green Internet of Things (GIoT): It is a version of the IoT that involves energy efficient features and environmental conservation activities. It ensures greener and healthier environment. First, the GIoT aims at reducing energy waste and minimizing energy consumption (Rafeeq, 2019). The GIoT technologies are also employed to significantly decrease the emissions of toxic pollutants such as carbon dioxide ($CO_2$)) and the greenhouse effect caused by the IoT systems (Albreem et al., 2021).

In this section, the most popular application domains of the IoT were presented. Their important aspects, application domains, target environments, target customers and their most important objectives are shown in Table 6.

The common challenges and application issues for the different versions of the IoT can be summarized as follows:

- Management of large amounts of generated and communicated data by collecting, analyzing and transmitting them.
- Security and confidentiality of the transmitted and stored data: consist in protecting IoT environments against possible threats.
- Considering the heterogeneity of the used technologies: The hardware and software installations implemented must guarantee compatibility and interoperability between the many used technologies.
- The scalability and connectivity of the IoT networks: are ensured by implementing scalable solutions to meet the future needs of networks where the number of connected objects is growing exponentially.
- Optimization of energy consumption: The deployed solutions and the used technologies must consume these resources in an optimized manner.

## 4  Conclusion

The success and benefits of the IoT lead to its rapid emergence in our domestic and professional life. Therefore, research works and technological innovations has recently evolved in order to address the different application domains. In fact, this paper presented a review describing the communication technologies used in the IoT networks. These technologies and their common use cases were compared. Subsequently, a variety of IoT application domains were listed. Moreover, each IoT application domain was depicted and examples of its use cases as well as its common challenges and problems were illustrated. Moreover, some recent research studies dealing with the IoT were summarized. We noticed that these application areas have the same challenges which consist in the management of Big Data (collection, transmission and analysis) as well as the security and confidentiality of these data and the interoperability between the heterogeneous used technologies. In this context, numerous intelligent methods can be applied to solve these problems. In fact, numerous intelligent and hybrid artificial intelligence-based methods were introduced and investigated the resolution of engineering real-world problems, especially IoT applications and issues (Mnasri et al., 2017) (Mouhammd et al., 2015) (MNASRI et al., 2020) (Hassanat, 2018) (Tarawneh et al., 2018) (MNASRI et al., 2018) (Hassanat et al., 2015). Most of these studies confirmed the complexity of IoT problems and highlighted the need for innovative methods to resolve IoT related challenges. As a continuation of this work, we will focus on the challenges and the unsolved problems of the IoT in order to explain them and propose promising ideas to find solutions.